\documentclass[conference]{IEEEtran}
\IEEEoverridecommandlockouts

\usepackage{commath}
\usepackage{cite}
\usepackage{subcaption}
\usepackage{amsthm}
\usepackage{amssymb}
\usepackage{amsmath}
\usepackage{array}
\usepackage{url}
 \usepackage{balance}
 \usepackage{multirow}
 \usepackage{booktabs}
 \usepackage{algorithm}
  \usepackage{fancyhdr}
 \usepackage{mathtools}
 \usepackage{placeins}
 \usepackage{bbm}
  \usepackage[font=footnotesize]{caption}
\usepackage{amsmath,amssymb,amsfonts}
\usepackage{algorithmic}
\usepackage{graphicx}
\usepackage{textcomp}
\usepackage{soul}
\def\Pr{\mathbb{P}}
\def\Ex{\mathbb{E}}
\def\peq#1{\stackrel{\text{\scriptsize(#1)}}{=}}

\newcounter{MYtempeqncnt}

\begin{document}
\title{A Tunable Base Station Cooperation Scheme\\ 
for Poisson Cellular Networks}

\author{\IEEEauthorblockN{Ke Feng}
\IEEEauthorblockA{Department of Electrical Engineering
\\University of Notre Dame}
\and
\IEEEauthorblockN{Martin Haenggi}
\IEEEauthorblockA{Department of Electrical Engineering\\University of Notre Dame\\}}

\maketitle
\IEEEpeerreviewmaketitle
\
\begin{abstract}
    We propose a tunable location-dependent base station (BS) cooperation scheme by partitioning the plane into three regions: the cell centers, cell edges and cell corners. The area fraction of each region is tuned by the cooperation level $\gamma$ ranging from 0 to 1. Depending on the region a user resides in, he/she receives no cooperation, two-BS cooperation or three-BS cooperation. Here, we use a Poisson point process (PPP) to model BS locations and study a non-coherent joint transmission scheme, $\textit{i.e.}$, selected BSs jointly serve one user in the absence of channel state information (CSI). For the proposed scheme, we examine its performance as a function of the cooperation level using tools from stochastic geometry. We derive an analytical expression for the \textit{signal-to-interference ratio} (SIR) distribution and its approximation based on the asymptotic SIR gain, along with the characterization of the normalized spectral efficiency per BS. Our result suggests that the proposed scheme with a moderate cooperation level can improve the SIR performance while maintaining the normalized spectral efficiency.
\end{abstract}

\section{Introduction}
 
The link quality in a cellular network strongly depends on the location of the users relative to the serving and interfering BSs. Specifically, in dense networks where interference is the limiting factor, the \textit{signal-to-interference ratio} (SIR) for
users distant from the serving BS is on average lower than that near the serving BS. Such a variation induces unfairness and harms the performance of users near the cell boundary \cite{gesbert2010multi}. BS cooperation is one of the methods to ameliorate the problem. By allowing selected BSs to jointly serve one or more users, the interference originating from nearby BSs can be turned into useful signals. Nonetheless, practical BS cooperation schemes need to be evaluated under the constraint of limited time-frequency resource blocks (RBs) at each BS. If the number of cooperating BSs per user keeps increasing, the gain will gradually diminish because distant BSs have little impact on the SIR, and the overall throughput will be reduced. In this respect, it is thus crucial to devise a location-dependent cooperation scheme and limit the number of cooperating BSs per user.

BS cooperation schemes vary based on how to group cooperating BSs, how BSs jointly serve users and whether they are adaptive to the user location.

\cite{baccelli2015stochastic} proposes a pairwise BS cooperation scheme where users within the cooperation region can be served by the two nearest BSs using coherent multi-user joint transmission. The transmission scheme, however, relies on precise channel state information (CSI) and intensive computation. 
A transmission scheme that is less sensitive to channel estimation and backhaul capacity is analyzed in \cite{nigam2014coordinated}, 
where the authors study a single-user joint transmission scheme in heterogeneous networks. Selected BSs non-coherently transmit the same desired symbol to serve a user.
While the cooperation scheme is not location-dependent, two types of users are studied, namely the general user and the worst-case user (users located at the Voronoi vertices). It is shown that the cooperation scheme benefits the worst-case user more significantly than the general user in terms of the SIR. In our paper, location-dependent BS cooperation will be studied to account for such a difference and the two types of users will be generalized to three types of users. Further,  \cite{garcia2014coordinated} proposes
a user-centric method of clustering BSs to maximize each user's normalized spectral efficiency, raising the importance of evaluating BS cooperation schemes in terms of the number of serving BSs per user. The evaluation of the normalized spectral efficiency as a function of the cooperation level will be included in this paper.

Here, our focus is the single-user non-coherent joint transmission scenario as in \cite{nigam2014coordinated}. We propose a cooperation scheme that favors users relatively distant from their serving BS. We first offer a crisp mathematical definition of the cell center, the cell edge and the cell corner. The division is adjusted by the cooperation level $\gamma$ that ranges from 0 to 1. Users in the cell center are served by only the nearest BS, while users in the cell edge and the cell corner are served by the two and the three nearest BSs, respectively. 
 
 We evaluate the performance of the proposed scheme by two metrics: the SIR distribution and the spectral efficiency normalized by the number of cooperating BSs. The former characterizes the typical link quality and the latter characterizes the overall throughput. Exact analytical expressions of the metrics are given, followed by an approximation based on the asymptotic SIR gain\cite{haenggi2014mean}. It is found in \cite{haenggi2014mean} that with the same diversity order, the SIR gain between different schemes can be captured using the horizontal gap given by the ratio of the \textit{mean interference-to-signal ratio} (MISR). Here, we study the analytical SIR gain between the standard Poisson point process (PPP) without cooperation and our scheme. We show that with the increase of the cooperation level, the SIR gain amounts more slowly and essentially saturates. The relative ditance process introduced in \cite{ganti2016asymptotics} is used to obtain this result.  

\section{Model}
\subsection{Geometric Division}

For a cooperation level $\gamma$ ($0\leq\gamma\leq1$) we partition the plane into three disjoint regions. Letting $\rho=1-\gamma$ we define
\begin{equation}
\begin{split}
    \label{eq:division}
&\mathcal{C}_1\triangleq\{x\in\mathbb{R}^2\colon \norm{x-\mathrm{NP}_1(x)}\leq \rho\norm{ {x-\mathrm{NP}_2(x)}}\},\\
&\mathcal{C}_2\triangleq\{x\in\mathbb{R}^2\colon\rho \norm{x-\mathrm{NP}_2(x)}<\norm{x-\mathrm{NP}_1(x)},\\
&\qquad\quad\norm{x-\mathrm{NP}_1(x)}\leq \rho \norm{x-\mathrm{NP}_3(x)}\},\\
&\mathcal{C}_3\triangleq
\{x\in\mathbb{R}^2\colon \norm{x-\mathrm{NP}_1(x)}> \rho \norm{x-\mathrm{NP}_3(x)}\},
\end{split}
\end{equation}
where $\mathrm{NP}_{i}(x)$ is the $i$th nearest BS to $x$. Note that in a 2D Voronoi diagram, any location on a Voronoi edge is equidistant from its two nearest BSs, and any Voronoi vertice is equidistant from its three nearest BSs. Intuitively,
 $\mathcal{C}_1$ denotes the cell center region where users are only close to the nearest BS and relatively far from other BSs, $\mathcal{C}_2$ denotes the cell edge region where users are relatively close to the two nearest BSs, and $\mathcal{C}_3$ denotes the cell corner region where users are relatively close to the three nearest BSs. 

Without cooperation, all users are connected to their nearest BS only. With location-dependent cooperation, a user residing in $\mathcal{C}_i$ is served by its $i$ nearest BSs. Note that $\gamma=0$ corresponds to the non-cooperation scheme and $\gamma=1$ corresponds to the full cooperation scheme where all users are connected to their three nearest BSs. 
\begin{figure}[t]
\begin{subfigure}[t]{0.49\linewidth}
\centering
\includegraphics[trim={98 20 95 15} ,clip,width=\linewidth]{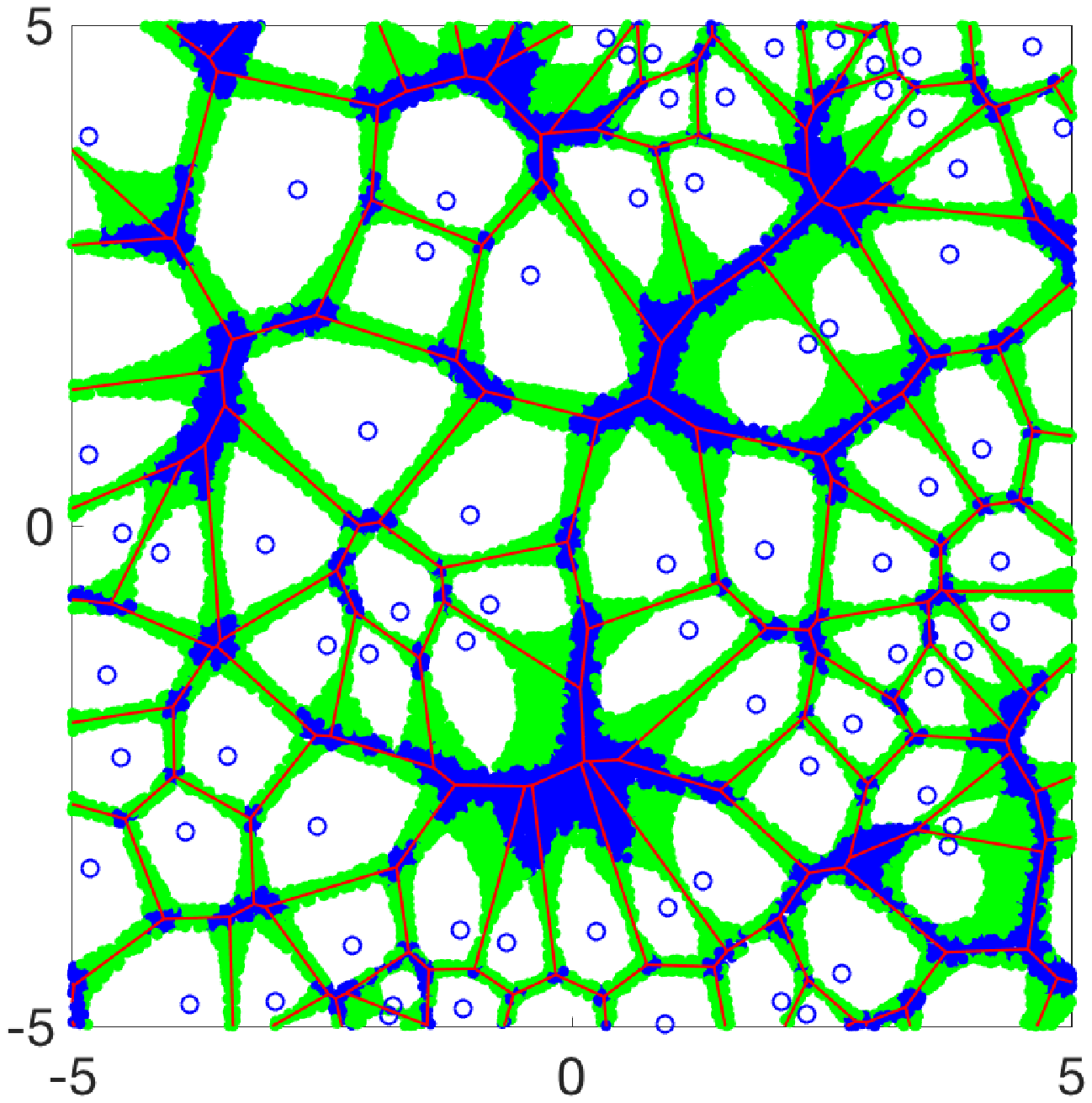}
\end{subfigure}
\hfill
\begin{subfigure}[t]{0.49\linewidth}
\centering
\includegraphics[trim={98 20 90 15} ,clip,width=\linewidth]{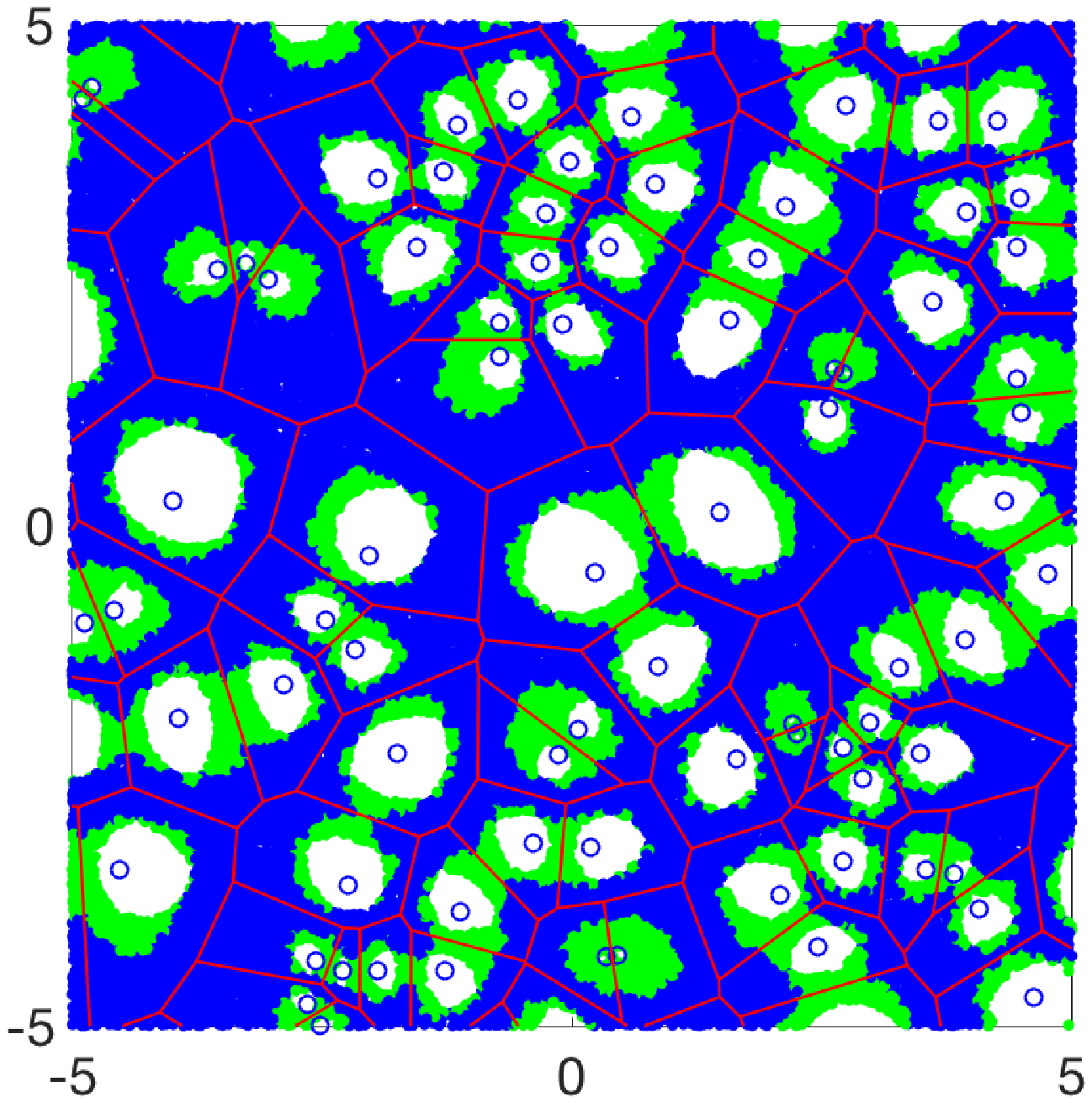}
\end{subfigure}%
\caption{Illustration of the cooperation region when $\gamma=0.2$ and $\gamma=0.5$. Blue circles denote points generated from a PPP of intensity 3. Red lines are the edges of the associated Voronoi cells. Blank, green and blue regions denote the non-cooperation region $\mathcal{C}_1$, the two-BS cooperation region $\mathcal{C}_2$ and the three-BS cooperation region $\mathcal{C}_3$ respectively. The area fractions are 0.64, 0.2304, 0.1296 for $\gamma =0.2$ and 0.25, 0.1875, 0.5625 for $\gamma =0.5$ obtained from (\ref{eq:frac}).}

\label{fig:1}
\end{figure}
\subsection{System Model}
Consider the downlink transmission in a cellular network with orthogonal RBs. Each BS transmits with unit power and is connected with its geometrical neighbors via backhaul links of sufficient capacity. We use a 2D PPP with intensity $\lambda$ to model BS locations, denoted as $\Phi$. Each BS/user is equipped with a single antenna. We focus on the typical user located at the origin $o$. The effect of Rayleigh fading from the point $x$ is denoted by $h_x$ and the path loss exponent by $\alpha$. The received signal at the typical user can be written as

\begin{equation}
\displaystyle\sum_{x\in \mathcal{C}}{\frac{h_xX}{{\|x\|}^{\alpha/2}}}+\displaystyle\sum_{x\in\Phi\setminus \mathcal{C}}{\frac{h_xX_x}{{\|x\|}^{\alpha/2}}}+Z,
\end{equation}
 where the first sum is the desired signal from the set of serving BS(s), denoted by $\mathcal{C}$ and the second sum is the interference from the other BSs. $X$ denotes the  channel input symbol sent by the serving BS(s) with zero mean and unit variance and is uncorrelated with other symbols $X_x$ sent by BSs not in $\mathcal{C}$. $Z$ is a zero mean complex Gaussian random variable with variance $\sigma^2$ modeling the background thermal noise.  
 
 Consider an interference-limited scenario, where the noise has little impact compared to the aggregated interference. The SIR at the typical user is
 \begin{equation}
\mathrm{SIR}=
\frac{{\left|\sum_{x\in \mathcal{C}}h_x{\|x\|}^{-\alpha/2}\right|}^2}{I}\nonumber
\end{equation}
with
\begin{equation}
I\triangleq\sum_{x\in\Phi\setminus \mathcal{C}}{|h_x|}^2{\|x\|}^{-\alpha}.\nonumber
\end{equation}
The correlation of the interfering items in $I$ resulting from interfering BSs' cooperation is ignored \cite{nigam2014coordinated}. 

 Further, let $r_i$ be the distance from the origin to its $i$-th nearest BS ( $r_{i}\leq r_{i+1}$ by definition). We define the distance point process $\Phi' = \{r_1, r_2, ...\}$  based on $\Phi$. The joint distribution of $r_1, r_2$ and $r_3$ is
\begin{equation}
    f_{r_1,r_2,r_3}(x,y,z)=(2\lambda\pi)^3xyz\exp{(-\lambda\pi z^2)}.
\end{equation}

The area fraction of each region depends on $\gamma$ as defined in (\ref{eq:division}) and is equal to the probability that the origin falls into each region \cite{6856159}:
\begin{equation}
\begin{split}
\label{eq:frac}
&\Pr(o \in \mathcal{C}_1)=(1-\gamma)^2\\
&\Pr(o \in \mathcal{C}_2)=\gamma(1-\gamma)^2(2-\gamma)\\
&\Pr(o \in \mathcal{C}_3)=\gamma^2(2-\gamma)^2.
\end{split}
\end{equation}
 An illustration of the partitioned plane when the cooperation level $\gamma=0.2$ and $\gamma=0.5$ is shown in Fig. \ref{fig:1}. The non-cooperation region corresponds to the locations near the center of each cell, the two-BS cooperation region follows the boundaries along the Voronoi cell edge, and the three-BS cooperation region closes around the Voronoi vertices.
\section{Performance Metrics}
\subsection{Success Probability}
For a given threshold $\theta$, the success probability with cooperation level $\gamma$ is defined as
\begin{equation}
\bar{F}_{\gamma}(\theta) \triangleq \Pr{(\mathrm{SIR}>\theta)},
\end{equation}
which is the complementary cumulative distribution function (ccdf) of the SIR.
\subsection{Asymptotic SIR Gain}
The SIR distribution of all but a few basic network models is complex or even intractable. In our scheme, the SIR distribution varies with the cooperation level $\gamma$. Hence, we simplify the success probability by calculating the asymptotic SIR gain between our scheme and the standard PPP without cooperation. It is shown in \cite{haenggi2014mean} that asymptotically,
\begin{equation}
\label{ASAPPP}
    \bar{F}_{\gamma}(\theta) \sim \bar{F}_{\mathrm{PPP}}(\theta/G),\quad\theta \to 0,
\end{equation}
where $\bar{F}_{\mathrm{PPP}}(\theta)$ denotes the ccdf of the SIR without cooperation. $G$ is reflected by the horizontal gap between SIR distributions and can be expressed as
\begin{equation}
G = \frac{\mathrm{MISR}_{\mathrm{PPP}}}{\mathrm{MISR}_{\gamma}},
\label{eq:MISR-ratio}
\end{equation}
where $\mathrm{MISR}_{\mathrm{PPP}}$ denotes the MISR of the PPP without cooperation and $\mathrm{MISR}_{\gamma}$ denotes the MISR of our cooperation scheme with cooperation level $\gamma$. 

The MISR is defined as \cite{haenggi2014mean}
\begin{equation}
    \mathrm{MISR}=\Ex{\Big(\frac{I}{\bar{S}}\Big)},
    \label{eq:MISR-I-S}
\end{equation}
where $I$ is the interference power as defined earlier, $S={{\left|\sum_{x\in \mathcal{C}}h_x{\|x\|}^{-\alpha/2}\right|}^2}$ and ${\bar{S}}=\Ex_{h}{(S)}$ is the signal power averaged over fading.


\subsection{Normalized Spectral Efficiency}
\begin{figure}[t]
\centering
\includegraphics[trim=0 0 0 0 ,clip,width=0.45\textwidth]{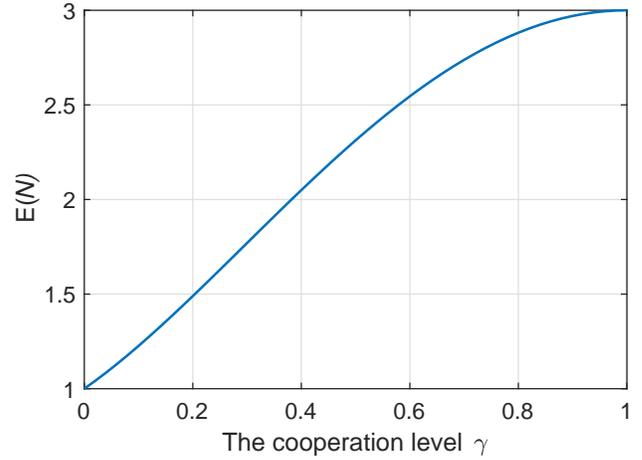} 
\caption{The mean number of BSs serving the typical user with respect to cooperation level $\gamma$.}
\label{fig:num_BS}
\end{figure}

We define the normalized spectral efficiency as 
\begin{equation}
C \triangleq {{N}^{-1}\log{(1+\mathrm{SIR})}},
\label{eq:capacity}
\end{equation} where $N$ denotes the number of serving BSs. The ergodic normalized spectral efficiency can be obtained by taking an expectation over (\ref{eq:capacity}). ${N}$ is a random variable that takes values from $\{1,2,3\}$ according to (\ref{eq:frac}) whose  mean is the mean number of BSs serving the typical user, $i.e.$,
\begin{equation}
 \Ex{{N}} = \gamma^4-4\gamma^3+3\gamma^2+2\gamma+1,
\end{equation}
which is shown in Fig. \ref{fig:num_BS}.
(\ref{eq:capacity}) captures the trade-off between the spectral efficiency and the overall throughput. Note that both ${N}$ and the SIR depend on which cooperation region the typical user falls in.

\section{Analytical and numerical results}

\subsection{Success Probability}
Without cooperation, the typical user is served by the nearest BS only. The success probability is given in \cite{6856159} as
\begin{equation}
\bar{F}_{\mathrm{PPP}}(\theta)=\frac{1}{_2F_1(1,-\delta;1-\delta;-\theta)},
\end{equation}
where $\delta\triangleq2/\alpha$ and $_2F_1(a,b;c;z)$ is the Gauss hypergeometric function. For $\alpha=4$ ($\delta=1/2$) we have $\bar{F}_{\mathrm{PPP}}(\theta)={(1+\sqrt{\theta}\arctan{\sqrt{\theta}})}^{-1}.$

In our cooperation scheme with cooperation level $\gamma$, we obtain the success probability by calculating the distribution of the SIR for the three regions. We obtain (\ref{eq:success_probability}) for the success probability when $\alpha=4$ ($\delta=1/2$) (see next page). The derivation of (\ref{eq:success_probability}) and the success probability for general $\alpha$ is provided in the appendix.

\begin{figure}[t]
\centering
\includegraphics[trim=0 0 0 0 ,clip,width=0.47\textwidth]{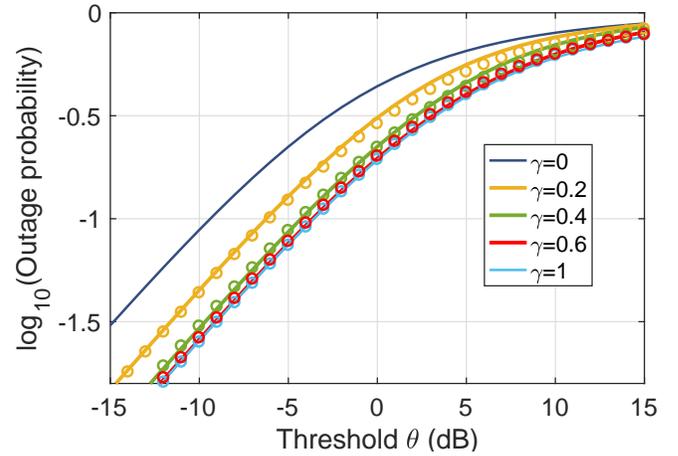}
\caption{The log outage probability with cooperation level $\gamma$ for $\alpha=4$ where the solid lines are plotted using the analytical form (\ref{eq:success_probability}) and circles are the asymptotic approximations using (\ref{ASAPPP}) and (\ref{eq:G}).}
\label{fig:shift}
\end{figure}
\begin{figure*}[t]
\normalsize
\setcounter{MYtempeqncnt}{\value{equation}}
\setcounter{equation}{\value{equation}}
\begin{align}
\label{eq:success_probability}
&\bar{F}_{\gamma}(\theta)= \sum_{i=1}^{3}\Pr{(\mathrm{SIR}>\theta, \mathcal{C}_i)}\\\nonumber
 & = \int_{0}^{\infty}{\int_{\frac{x}{\rho}}^{\infty}{(2\pi)^2xy\exp{(-\pi y^2)}\exp{\Big(-\pi\sqrt{{\theta}{x^{4}}}\tan^{-1}{{\sqrt{\theta{x^4/y^4}}}}\Big)}\frac{1}{{1+\theta x^4/y^{4}}}\mathrm{d}x\mathrm{d}y}}\\\nonumber
 &+\int_{0}^{\infty}{\int_{x}^{\frac{x}{\rho}}\int_{\frac{x}{\rho}}^{\infty}{(2\pi)^3xyz\exp{(-\pi z^2)}\exp{\Bigg(-\pi\sqrt{\frac{\theta}{x^{-4}+y^{-4}}}\tan^{-1}{\sqrt{\frac{\theta z^{-4}}{x^{-4}+y^{-4}}}}\Bigg)}\frac{1}{1+\theta z^{-4}/(x^{-4}+y^{-4})}\mathrm{d}x\mathrm{d}y}\mathrm{d}z}\\\nonumber
& +\int_{0}^{\infty}{\int_{x}^{\frac{x}{\rho}}\int_{y}^{\frac{x}{\rho}}{(2\pi)^3xyz\exp{(-\pi z^2)}\exp{\Bigg(-\pi\sqrt{\frac{\theta}{x^{-4}+y^{-4}+z^{-4}}}\tan^{-1}\sqrt{\frac{{\theta z^{-4}}}{x^{-4}+y^{-4}+z^{-4}}}\Bigg)}\mathrm{d}x\mathrm{d}y}\mathrm{d}z}.\\\nonumber
\end{align}
\setcounter{equation}{\value{MYtempeqncnt}+1}
\hrulefill
\vspace*{4pt}
\end{figure*}
\begin{figure*}[t]
\normalsize
\setcounter{MYtempeqncnt}{\value{equation}}
\setcounter{equation}{\value{equation}}
\begin{equation}
\label{eq:G}
    G=\frac{2}{{(\alpha+2)}\Ex{\Big((\frac{r_1}{r_2})^{\alpha}\mathbbm{1}_{\mathcal{C}_1}\Big)}+{(\alpha+4)}\Ex{\Big(\frac{(r_1/r_3)^{\alpha}}{1+(r_1/r_2)^{\alpha}}\mathbbm{1}_{\mathcal{C}_2}\Big)}+{6}\Ex{\Big(\frac{(r_1/r_3)^{\alpha}}{1+(r_1/r_2)^{\alpha}+(r_1/r_3)^{\alpha}}\mathbbm{1}_{\mathcal{C}_3}}\Big)}.
\end{equation}
\setcounter{equation}{\value{MYtempeqncnt}+1}
\hrulefill
\vspace*{4pt}
\end{figure*}

\subsection{Asymptotic Success Probability}
\textit{The Asymptotic Gain}:
 We obtain the asymptotic horizontal gap $G$ as (\ref{eq:G}) using the MISR of the PPP without cooperation and the MISR in our scheme with cooperation level $\gamma$. The former is \cite{haenggi2014mean}
\begin{equation}
    \mathrm{MISR}_{\mathrm{PPP}}=\frac{2}{\alpha-2},\nonumber
\end{equation} and the latter is calculated by applying the relative distance process introduced in \cite{ganti2016asymptotics}.
We determine $\mathrm{MISR_{\gamma}}$ by calculating it for the three regions and adding the results, $i.e.$,
\begin{equation}
    \mathrm{MISR_{\gamma}}= \mathrm{MISR}_{\mathcal{C}_1}+\mathrm{MISR}_{\mathcal{C}_2}+\mathrm{MISR}_{\mathcal{C}_3}\nonumber,
\end{equation}
where $\mathrm{MISR}_{\mathcal{C}_i}$ denotes the MISR calculated using (\ref{eq:MISR-I-S}) within ${\mathcal{C}_i}$. For ${\mathcal{C}_1}$, we have
\begin{align}
    \mathrm{MISR}_{{\mathcal{C}_1}} &=\sum_{i>1}\Ex{{\Big[\Big(\frac{r_1}{r_i}\Big)^{\alpha}}\mathbbm{1}_{\mathcal{C}_1}\Big]}\\\nonumber
     &\peq{a}\Ex{\Big[\Big(\frac{r_1}{r_2}\Big)^{\alpha}\mathbbm{1}_{\mathcal{C}_1}\Big]}{\sum_{i>1}\Ex\Big[\Big(\frac{r_2}{r_i}\Big)^{\alpha}\Big]},\nonumber
\end{align}
    where $\mathbbm{1}_{\mathcal{C}_i}$ is an indicator function that is one if the typical user falls into ${\mathcal{C}_i}$ and is zero otherwise. Step (a) follows from the fact that only the first term in $\mathrm{MISR}_{\mathcal{C}_1}$ is constrained by the cooperation region. It can be calculated using the joint distribution of $r_1$ and $r_2$ as
    \begin{equation}
    \Ex{\Big[\Big(\frac{r_1}{r_2}\Big)^{\alpha}\mathbbm{1}_{\mathcal{C}_1}\Big]}=\int_0^{\infty}\int_{\frac{x}{\rho}}^{\infty} f_{r_1,r_2}(x,y) \Big(\frac{r_1}{r_2}\Big)^{\alpha} \mathrm{d}y\mathrm{d}x\nonumber.
    \end{equation}
   The second term can be evaluated by considering the relative distance process \cite{ganti2016asymptotics} as 
    \begin{equation}
    \sum_{i>1}\Ex{\Big(\frac{r_2}{r_i}\Big)^{\alpha}}= 1+\frac{4}{\alpha-2}.\nonumber
    \end{equation}
    Similarly, we obtain the MISR in ${\mathcal{C}_2}$ and ${\mathcal{C}_3}$ as
\begin{align}
    \mathrm{MISR}_{\mathcal{C}_2} &=\sum_{i>2}\Ex{\Big[{\frac{r_i^{-\alpha}}{r_1^{-\alpha}+r_2^{-\alpha}}}\mathbbm{1}_{\mathcal{C}_2}\Big]}\\\nonumber
     &=\Ex{\Big[\frac{(r_1/r_3)^{\alpha}}{1+(r_1/r_2)^{\alpha}}\mathbbm{1}_{\mathcal{C}_2}\Big]}\sum_{i>2}\Ex{\Big[\Big(\frac{r_3}{r_i}\Big)^{\alpha}\Big]}\nonumber,
\end{align}
\begin{equation}
\sum_{i>2}\Ex{\Big(\frac{r_3}{r_i}\Big)^{\alpha}}=1+\frac{6}{\alpha-2},\nonumber
\end{equation}
and
\begin{align}
    \mathrm{MISR}_{\mathcal{C}_3} &=\sum_{i>3}\Ex{\Big[{\frac{r_i^{-\alpha}}{r_1^{-\alpha}+r_2^{-\alpha}+r_3^{-\alpha}}}\mathbbm{1}_{\mathcal{C}_3}\Big]}\\\nonumber
     &=\Ex{\Big[\frac{(r_1/r_3)^{\alpha}}{1+(r_1/r_2)^{\alpha}+(r_1/r_3)^{\alpha}}\mathbbm{1}_{\mathcal{C}_3}\Big]}\sum_{i>3}\Ex{\Big[\Big(\frac{r_3}{r_i}\Big)^{\alpha}\Big]},\nonumber
\end{align}

\begin{equation}
    \sum_{i>3}\Ex{\Big(\frac{r_3}{r_i}\Big)^{\alpha}}=\frac{6}{\alpha-2}.\nonumber
\end{equation}
Now, using (\ref{eq:MISR-ratio}) we obtain the expression for $G$ as in (\ref{eq:G}). As shown in Fig. \ref{fig:shift}, the approximation of the outage probability (defined as the cdf of the SIR) based on the horizontal gap is very accurate compared to the exact analytical result (\ref{eq:success_probability}).

Approximating the analytical $G$ in (\ref{eq:G}) using basic fitting (in dB) we get
\begin{equation}
\label{eq:G_fit}
G_{\mathrm{fit}} = a\tanh(b\gamma),\quad\mathrm{(in~dB)}
\end{equation}
where $a= 5.865$ and $b=3.234$. 
Using (\ref{eq:G_fit}) and (\ref{ASAPPP}) we obtain the asymptotic form of the success probability as a function of $\theta$ and $\gamma$. 

\begin{figure}[t]
\centering
\includegraphics[trim=0 0 0 0 ,clip,width=0.45\textwidth]{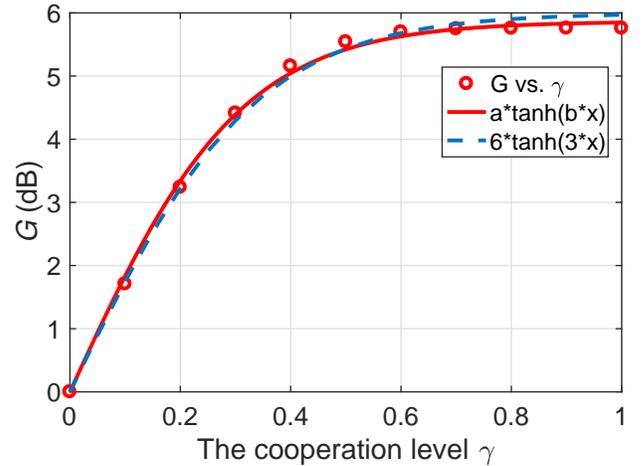}
\caption{Basic fitting of the horizontal gap $G$ (in dB) for $\alpha=4$ using $\tanh$ function with $a=5.865$ and $b=3.234$. The function with $a=6$ and $b=3$ also shows good fit with further simplicity. Red dots are plotted using the analytical expression (\ref{eq:G}).}
\label{fig:fit2}
\end{figure}

A further simplification of the approximation is\begin{equation}
\label{eq:G_fit2}
\tilde{G}_{\mathrm{fit}} = 6\tanh(3\gamma),\quad\mathrm{(in~dB)}
\end{equation}
which is surprisingly simple and still quite accurate as shown in Fig. \ref{fig:fit2} .
\begin{figure}[t]
\centering
\includegraphics[trim=0 0 0 0 ,clip,width=0.45\textwidth]{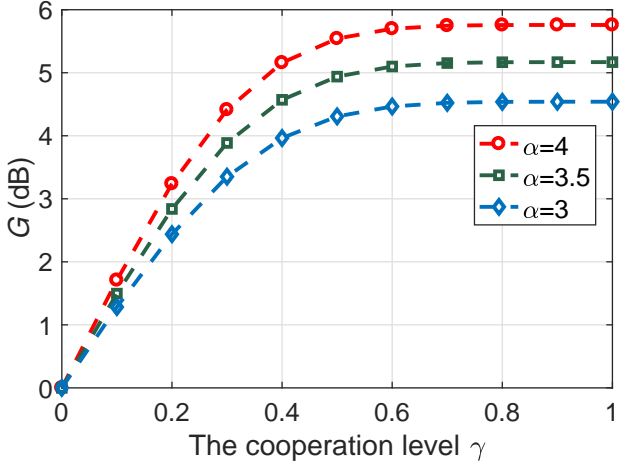}
\caption{Horizontal shift (in dB) using (\ref{eq:G}) for different $\alpha$.}
\label{fig:G}
\end{figure}

\textit{G \textit{vs.} $\alpha$}:
The comparison of $G$ for different $\alpha$ is shown in Fig. \ref{fig:G}. As the path loss exponent $\alpha$ increases, the horizontal gain increases also. Note that the horizontal gain, regardless of $\alpha$, is increasing almost linearly at first, and amounts very slowly after $\gamma=0.6$. It suggests that higher cooperation levels beyond the threshold essentially offer no further SIR gain, which means the overall throughput decreases since $N$ increases with $\gamma$.
\subsection{Ergodic Normalized Spectral Efficiency}
\begin{figure}[t]
\centerline{\includegraphics[trim=0 0 0 0 ,clip,width=0.47\textwidth]{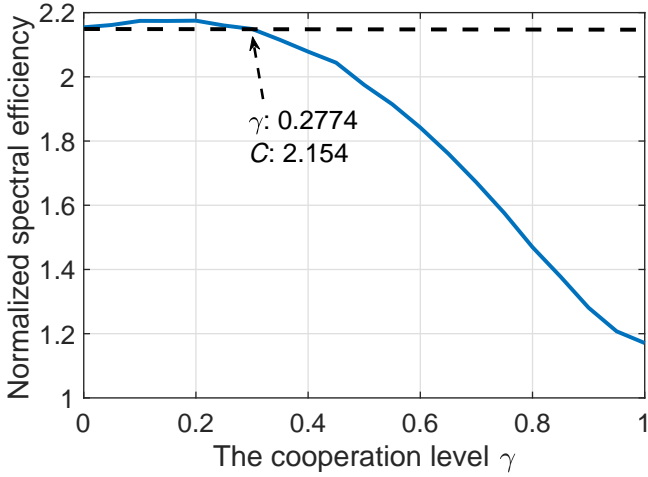} }
\caption{The simulation result of the ergodic normalized spectral efficiency for $\alpha=4$.}
\label{fig:simu-capacity1}
\end{figure}

 As shown in the simulation results in Fig. \ref{fig:simu-capacity1}, the ergodic normalized spectral efficiency increases slightly and then decreases with the increase of $\gamma$ ($i.e.$, the expansion of the cooperation region $\mathcal{C}_2\cup\mathcal{C}_3$). Observe that the same normalized spectral efficiency is guaranteed when $\gamma=0$ and $\gamma\approx0.28$, which gives the range of cooperation levels that improve the typical link quality without lowering the overall throughput.

\section{Conclusions}
  The tunable BS cooperation scheme put forth in this paper offers cooperation that is adaptive to a user's position. It is shown that with the increase of the cooperation level, the SIR gain saturates quickly after $\gamma=0.6$, while the number of serving BSs keeps increasing. The normalized spectral efficiency further validates that moderate cooperation is optimal under limited BS resources. In essence, the proposed BS cooperation scheme not only adaptively allocates resources in the network to boost the signal strength and mitigates the interference but also compensates unfairness, by allocating more resources to users distant from the nearest BS.
  \section*{Acknowledgment}
The support of the U.S.~NSF (grant CCF 1525904) is gratefully acknowledged.
\appendix
\subsection{General $\alpha$}
The success probability with $\gamma$ can be written as
\begin{align}
&\Pr{(\mathrm{SIR}>\theta)}=\sum_{i=1}^{3}\Pr{(\mathrm{SIR}>\theta, \mathcal{C}_i)},
\label{eq:success-probability-general-alpha}
\end{align}
where $\Pr{(\mathrm{SIR}>\theta, \mathcal{C}_i)}$ is the ccdf of the SIR if the typical user falls in cooperation region $\mathcal{C}_i$. For example,
\begin{align}
    \Pr{(\mathrm{SIR}>\theta, \mathcal{C}_1)}&=\Pr{\Bigg(g_1r_1^{-\alpha}>\theta\bigg(\sum_{i\neq 1} g_ir_i^{-\alpha}\bigg),\mathcal{C}_1\Bigg)}\\\nonumber
    &\peq{a} \Pr{\left(g_1>\theta Ir_1^{\alpha},\mathcal{C}_1\right)}\\\nonumber
    &\peq{b}\int_{\mathcal{C}_1}{\mathcal{L}^{(1)}_{I}{\left({\theta }{r_1^{\alpha}}\right)}}f_{r_1,r_2}(x,y)\mathrm{d}x\mathrm{d}y\nonumber
\end{align}
where (a) follows from $g_i \sim \exp(1)$ and $I=\sum_{i\neq 1} g_ir_i^{-\alpha}$, and (b) follows from the Laplace transform of the interference $I$, $i.e.,$ ${\mathcal{L}^{(1)}_{I}{(s)}}$, evaluated at $s={\theta}{r_1^{\alpha}}$.
${\mathcal{L}^{(1)}_{I}{(s)}}$ can be written as
\begin{align*}
    &\mathcal{L}^{(1)}_{I}(s)\\\nonumber
    &\triangleq\Ex\left(e^{-s\sum_{i=2}^{\infty}{g_ir_i^{-\alpha}}} \right)\\\nonumber
    &=\Ex_{\Phi'}\left(\prod {\frac{1}{1+sr_i^{-\alpha}}}\right)\\\nonumber
    &\peq{a} \exp{\left(-\int_{{r_2}}^{\infty} \left[1-\frac{1}{1+s x^{-\alpha}}\right]2\lambda\pi x\mathrm{d}x\right)}\frac{1}{1+sr_2^{-\alpha}}\\\nonumber
    &\peq{b}
    \exp{\left(-2\lambda \pi s^{\frac{2}{\alpha}}\int_{{r_2s^{-\frac{1}{\alpha}}}}^{\infty} \frac{t}{1+t^{\alpha}}\mathrm{d}t\right)}\frac{1}{1+sr_2^{-\alpha}}\\\nonumber
    &\peq{c}
    \exp{\left(-2\lambda \pi s^{\delta}F(r_2s^{-\frac{\delta}{2}})\right)}\frac{1}{1+sr_2^{-2/\delta}},\nonumber
\end{align*}
where $\delta=2/\alpha$, $\Phi'$ is the distance point process as defined before,
(a) is due to the probability generation functional of the PPP \cite{haenggi2012stochastic}, (b) follows from the substitution $x = s^{\frac{1}{\alpha}}t$, and (c) follows from the definition  $F(x)\triangleq\int_{x}^{\infty}\frac{t}{1+t^{\alpha}}\mathrm{d}t$. The integral can be expressed in terms of the Gauss hypergeometric function 
\begin{equation}
    F(x)={\frac{x^2}{(2/\delta-2)(1+x^{2/\delta})}} {_2F_1\left(1,1;2-\delta;\frac{1}{1+x^{2/\delta}}\right)},
\end{equation} and can be easily evaluated numerically.

Similarly, the success probability of the typical user in the cell-edge region can be written as
\begin{equation}
\begin{split}
    &\Pr{(\mathrm{SIR}>\theta, \mathcal{C}_2)}\\
    &= \Pr{\Bigg(g_1r_1^{-\alpha}+g_2r_2^{-\alpha}>\theta\bigg(\sum_{i\neq 1,2} g_ir_i^{-\alpha}\bigg),\mathcal{C}_2\Bigg)}\\\nonumber
\end{split}
\end{equation}
and
\begin{equation}
\begin{split}
    &\mathcal{L}^{(2)}_{I}(s)\\
    &=\mathrm{E}\left(e^{-s\sum_{i=3}^{\infty}{g_ir_i^{-\alpha}}}\right)\\
    &=\mathrm{E}_{\Phi'}\left(\prod {\frac{1}{1+sr_i^{-\alpha}}}\right)\\
    &= \exp{\left(-\int_{r_3}^{\infty} \left[1-\frac{1}{1+s x^{-\alpha}}\right]2\lambda\pi x\mathrm{d}x\right)}\frac{1}{1+sr_3^{-\alpha}}\\
    &=
    \exp{\left(-2\lambda \pi s^{\delta}F(r_3s^{-\frac{\delta}{2}})\right)}\frac{1}{1+sr_3^{-2/\delta}}
    .
\end{split}
\nonumber
\end{equation}

The success probability of the typical user in the cell-corner region can be written as
\begin{equation}
\begin{split}
    &\Pr{(\mathrm{SIR}>\theta, \mathcal{C}_3)}\\
    &= \Pr{\Bigg(g_1r_1^{-\alpha}+g_2r_2^{-\alpha}+g_3r_3^{-\alpha}>\theta\bigg(\sum_{i\neq 1,2,3} g_ir_i^{-\alpha}\bigg),\mathcal{C}_3\Bigg)}\\\nonumber
\end{split}
\end{equation}
and
\begin{equation}
\begin{split}
    \mathcal{L}^{(3)}_{I}(s)&=\mathrm{E}\left(e^{-s\sum_{i=4}^{\infty}{g_ir_i^{-\alpha}}}\right)\\
    &=\mathrm{E}_{\Phi'}\left(\prod {\frac{1}{1+sr_i^{-\alpha}}}\right)\\
    &=\exp{\left(-\int_{r_3}^{\infty} \left[1-\frac{1}{1+s x^{-\alpha}}\right]2\lambda\pi x\mathrm{d}x\right)}\\
    &=
    \exp{\left(-2\lambda \pi s^{\delta}F\big(r_3s^{-\frac{\delta}{2}}\big)\right)}.
\end{split}
\nonumber
\end{equation}
By summing up the three terms we obtain the success probability for general $\alpha$ with cooperation level $\gamma$.
\subsection{Special Case: $\alpha=4$}
For $\alpha=4$, we can simplify the above conditional Laplace transforms of the interference to closed forms:
\begin{align}\nonumber
&\mathcal{L}^{(1)}_{I}(s)=\exp{(-\lambda\pi\sqrt{s}\arctan{{\sqrt{s}}{r_2^{-2}}})}\frac{1}{1+sr_2^{-4}},\\\nonumber
&\mathcal{L}^{(2)}_{I}(s)=\exp{(-\lambda\pi\sqrt{s}\arctan{{\sqrt{s}}{r_3^{-2}}})}\frac{1}{1+sr^{-4}_3},\\\nonumber
&\mathcal{L}^{(3)}_{I}(s)=\exp{(-\lambda\pi\sqrt{s}\arctan{{\sqrt{s}}{r^{-2}_3}})}.\nonumber
\end{align}

By plugging the above equations into (\ref{eq:success-probability-general-alpha}) we obtain the success probability for $\alpha=4$ in (\ref{eq:success_probability}). 

 \bibliographystyle{IEEEtran}


\end{document}